\documentclass[a4paper,11pt]{article}
\pdfoutput=1 

\usepackage{jinstpub} 

\title{\boldmath Characterization of a defective PbWO$_4$ crystal cut along the a-c crystallographic plane: structural assessment and a novel photoelastic stress analysis}

\author[a]{L. Montalto,}
\author[b]{P.P. Natali,}
\author[b]{F. Dav\'{\i}}
\author[c]{P. Mengucci}
\author[a]{N. Paone}
\author[c,1]{D. Rinaldi\note{Corresponding author.}}

\affiliation[a]{DIISM,}
\affiliation[b]{DICEA,}
\affiliation[c]{SIMAU,\\Universit\'a Politecnica delle Marche, via Brecce Bianche 12, 60131 Ancona, Italy}

\emailAdd{d.rinaldi@univpm.it}

\abstract{Among scintillators, the PWO is one of the most widely used, for instance in CMS calorimeter at CERN and PANDA project. Crystallographic structure and chemical composition as well as residual stress condition, are indicators of homogeneity and good quality of the crystal. In this paper, structural characterization of a defective PbWO$_4$ (PWO) crystal has been performed by X-ray Diffraction (XRD), Energy Dispersive Spectroscopy (EDS) and Photoelasticity in the unusual $(a, c)$ crystallographic plane. XRD and EDS analysis have been used to investigate crystallographic orientation and chemical composition, while stress distribution, which indicates macroscopic inhomogeneities and defects, has been obtained by photoelastic approaches, in Conoscopic and Sphenoscopic configuration. Since the sample is cut along the $(a, c)$ crystallographic plane, a new method is proposed for the interpretation of the fringe pattern. 
The structural analysis has detected odds from the nominal lattice dimension, which can be attributed to the strong presence of Pb and W. A strong inhomogeneity over the crystal sample has been revealed by the photoelastic inspection. The results give reliability to the proposed procedure which is exploitable in crystals with other structures.
}


\begin{document}
\maketitle
\flushbottom

\section{Introduction}

Application areas of scintillating crystals span from high energy physics to security, to geological prospections and medical applications \cite{1, 2}. PWO, as a scintillating material, is continuing getting attention from the high energy physics community \cite{3, 4}, since the time of CMS \cite{5} and, also, today with the PANDA \cite{4}-\cite{6} project, thanks to its intrinsic characteristics. Owing to its own scintillating properties, the tetragonal PbWO$_4$ (PWO) crystals are among the widely used and studied \cite{2}, \cite{4}, \cite{7}-\cite{9}. The fields themselves prompt the efforts in theoretical, experimental and technological studies of these substances and the associated production processes. In addition to purely theoretical aspects, from the technological point of view, a high-quality mass production requires to deepen the knowledge of crystal structure as well as of mechanical and optical behavior of crystals, in order to address growth process technologies. From the industrial point of view, the development of procedures and reliable methods of inspection and characterization is mandatory for a more and more efficient production, improving the process control, the crystal quality and reducing the costs. In order to achieve these goals, the development of experimental techniques and theoretical models, capable of improving knowledge of crystal behavior as well as their characteristics and structure, is necessary \cite{8}, \cite{10, 11}.
In most cases, quality control procedures involve identification and quantification of defects present in the crystal sample. For this purpose, methods aimed at the study and the inspection of matter like X-ray diffraction (XRD), scanning electron microscopy (SEM) and microanalysis (EDS) allow for a quantitative characterization of crystalline structure, composition and lattice orientation. Fast, non-destructive and reliable quality control techniques of birefringent material are based on optical crystallography \cite{12} and photoelasticity \cite{13, 14}. These techniques are aimed at the study and investigation of the interference isochromate and isogyre fringes, generated by birefringent effects. The fringe patterns depend by the crystal structure \cite{15, 16}, stress state (applied and/or residual) and defectiveness of the sample.
Photoelastic studies, carried out by the implementation of reliable polariscopes \cite{17, 18}, are fundamental in the comprehension of defect dynamics and crystal reliability. 
For PWO and similar materials, photoelastic analysis is quite well established in case of samples observed in the plane normal to the optic axis \cite{18}-\cite{21}. On the contrary, although samples cut parallel to the optic axis have been already observed \cite{12}, \cite{22}, fringe patterns in this configuration have not been completely analysed yet (to our best knowledge), despite their importance for both crystal theoretical knowledge and assessment of production processes reliability and efficiency.
In this paper, we report results of structural characterization of a quasi-cylindrical PWO, performed by X-ray diffraction (XRD), scanning electron microscopy (SEM) and microanalysis (EDS) as well as photoelastic investigations, carried out by Conoscopy and Sphenoscopy \cite{23}. The sample is cut with the c axis parallel to the observation plane, therefore a model, which links easy measurable quantities of the fringe pattern to the sample stress condition, is proposed. The model for the data interpretation, which is an innovative contribution of the work, is based on studies already reported in literature \cite{14, 15}, \cite{24}-\cite{26} and allows for information on both stress and the piezo-optic tensor $\Pi$ components \cite{27}-\cite{29}. 
The obtained results put in evidence a great inhomogeneity of the sample, which are well in accord with the clear defectiveness of the PWO sample and underline the reliability of the photoelastic methods and the related model.

\section{Experiments}

\subsection{Sample description}

The quasi-cylindrical single crystal of lead tungsten oxide has a nominal composition PbWO$_4$ (stolzite) and it has been grown by the Czochralski method. The PbWO$_4$ (PWO) has a body-centered tetragonal crystallographic structure with lattice parameters $a=0.54619$ nm and $c=1.2049$ nm (ICDD card n.19-708). PWO is a high-density material with  $\rho= 8.26$ g/cm$^3$. This crystal is uniaxial negative ($n_{o}>n_{e}$), with extraordinary and ordinary refraction indices $n_{e}=2.163$ and $n_{o}=2.234$, for the visible radiation with $\lambda= 632.8$ nm \cite{11}.
The ingot grown with its axis normal to the crystallographic $c-$axis has been cut into slices, so that the investigated surface should be parallel to the $c-$axis of the crystal.

\begin{figure}[htbp]
\centering 
\includegraphics{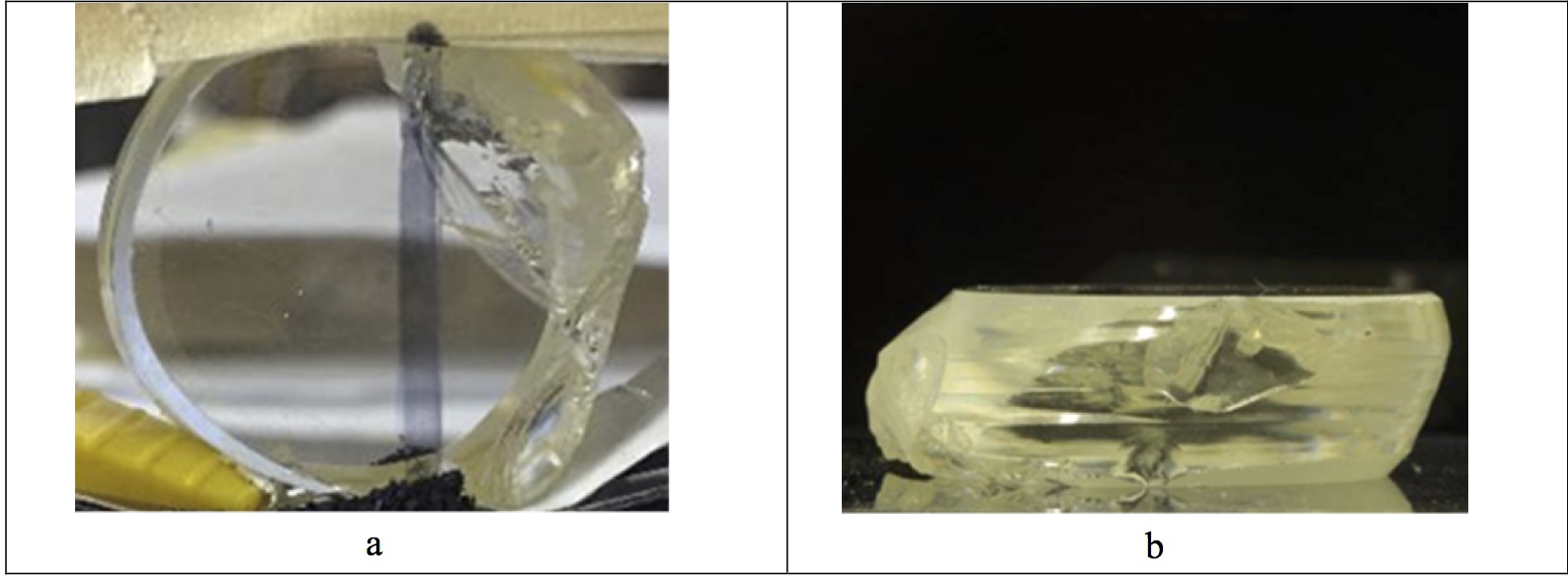}
\caption{\label{fig1} Pictures of the defective sample. In the Figure (a), the black line traced on the surface indicates the direction of the $c$ crystallographic axis; scretches and chipping occured during the machining are clearly visible on the surface. By the Figure (b), the distorted cylindrical shape is evident. Both the issue can be due to the rise of residual stress and defects during the growth.}
\end{figure}

In Figure~\ref{fig1}, the pictures of the PWO slice under inspection are reported; the sample presents some defects clearly visible by the front and the side picture. 

Visible scratches and chipping on the top right of the Figure~\ref{fig1}a are defects occurred during the machining while, the sliding effect, which induced the quasi-cylindrical geometry, is due to the grown procedures.
Both the issues can be referable to the rising and the consolidation of a residual stress distribution during the growth. This has led to a non-homogeneous state.

\subsection{Structural characterization}

The investigation of the structure of this PWO sample has been carried out by X-ray Diffraction (XRD), Scanning Electron Microscopy (SEM) and Energy Dispersive Spectroscopy (EDS).

\begin{figure}[htbp]
\centering 
\includegraphics{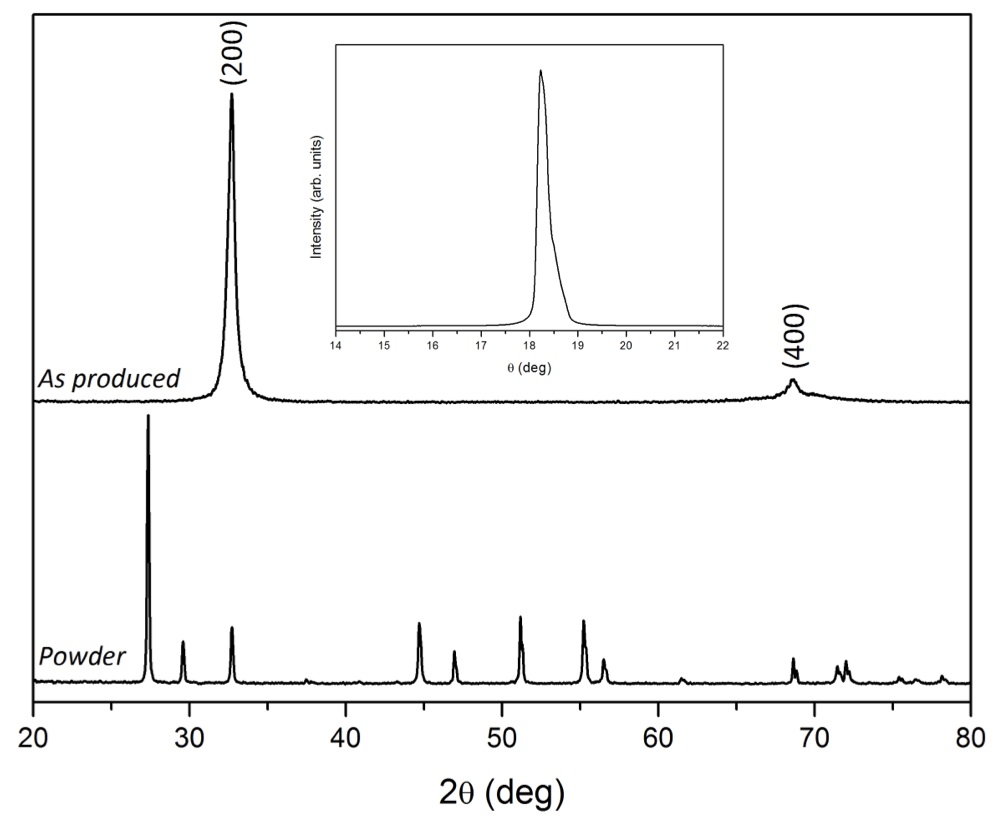}
\caption{\label{fig2} XRD patterns of PWO crystal in as produced (AP) condition and after grinding to powder. Inset shows the rocking curve obtained from the AP sample.}
\end{figure}

X-ray diffraction (XRD) measurements have been performed by a Bruker D8 Advance diffractometer operating at V= 40 kV and I= 40 mA, with Cu-K($\alpha$) radiation. A $\theta\div 2\theta$ geometry has been used to investigate growth orientation of the crystal while a rocking curve (RC) in the angular range $\theta=14^{\circ}\div 22^{\circ}$ has been carried out to put into evidence possible preferential orientations. Furthermore, in order to measure the lattice parameters of the crystal, an XRD powder pattern in the angular range $2\theta=20^{\circ}\div 80^{\circ}$ has been obtained from a splinter of the sample reduced to powder by manually grinding into a mortar. Peak analysis has been performed using the Origin software [www.origiblab.com]. The quantitative results of peak analysis are reported as provided by Origin.
Scanning electron microscopy (SEM) observations have been carried out by a Zeiss Supra 40 field emission microscope equipped with a Bruker Z200 energy dispersive (EDS) microanalysis.
The resulting patterns of the XRD analyses performed on the as-produced (AP) sample and on sample reduced to powder are reported in Figure~\ref{fig2}. Furthermore, the rocking curve obtained from the AP sample to check the degree of the preferential growth is shown in the inset of Figure~\ref{fig2}.

The XRD pattern of the AP sample (Figure~\ref{fig2}) clearly evidences a $(2,0,0)$ preferential growth. This means that the $(2,0,0)$ lattice planes of the PWO crystal grow parallel to the sample surface. The rocking curve reported in the inset of Figure~\ref{fig2} confirms the preferential orientation growth. Quantitative analysis of the $(2,0,0)$ reflection in Figure~\ref{fig2} (AP sample) allows estimating the exact peak position at $2\theta=32.686\pm0.001^{\circ}$ with a full width at half maximum (FWHM) of $0.507\pm0.003^{\circ}$. Compared to the nominal value ($2\theta=32.754^{\circ}$) of the $(2,0,0)$) reflection for the PbWO$_4$ (Stolzite) compound reported in the ICDD (International Centre for Diffraction Data) card n.19-708, the experimental value obtained in our case results to be only $0.2$\% lower.

The rocking curve in the inset in Figure~\ref{fig2} shows a peak located at $\theta=18.278\pm0.001^{\circ}$ with a FWHM of $0.236\pm0.003^{\circ}$. Considering the angular position of the $(2,0,0)$ peak in the XRD pattern of the AP sample, the rocking curve evidences a preferential growth of the $(2,0,0)$ lattice planes shifted of about $4^{\circ}$ from the expected position.

XRD pattern obtained from the sample, reduced to powder, allowed estimating the lattice parameters of the PWO crystal. Best fitting of peak position confirmed the tetragonal structure of the crystal and provided the following lattice parameters: $a=0.54676\pm0.00015$ nm and $c=1.2064\pm0.0004$ nm. These experimental values are about $0.1$\% higher than the nominal values reported in the ICDD card n.19-708 for PbWO$_4$ (Stolzite), suggesting a slightly larger crystallographic cell for our sample.

EDS analysis performed on large areas of the sample exclusively showed the presence of Pb, W and O without any further dopant or contaminant. Quantitative EDS provided the average content reported in Table~\ref{tab1}, where the nominal content of the PbWO$_4$ compound is also reported for comparison. Average values (AV) and corresponding standard deviations (SD) in Table~\ref{tab1} have been calculated from ten different EDS measurements on large areas of the AP sample.

\begin{table}[htbp]
\centering
\caption{\label{tab1} Experimental values of elements concentration in at.\% provided by EDS analysis of the AP sample. Average value (AV) and standard deviation (SD) have been calculated from ten different measurements. Nominal element content refers to the PbWO$_4$ compound.}
\smallskip
\begin{tabular}{|c|c|c|c|c|c|c|}
\hline
&Pb & &W& &O&\\
&(at \%)& &(at \%)& &(at \%)&\\
\hline
 &AV&SD &AV&SD&AV&SD\\
 \hline
AP sample &20.73&1.04 &21.97&1.21 &57.3&2.2\\
\hline
Nominal &16.7& &16.7& &66.7&\\
\hline
\end{tabular}

\end{table}

Results in Table~\ref{tab1} clearly show a higher content of Pb and W in the investigated sample with respect to the nominal composition of the PbWO$_4$ compound.

\subsection{Novel Photoelastic measurements in the $(a-c)$ plane}

\begin{figure}[htbp]
\centering 
\includegraphics{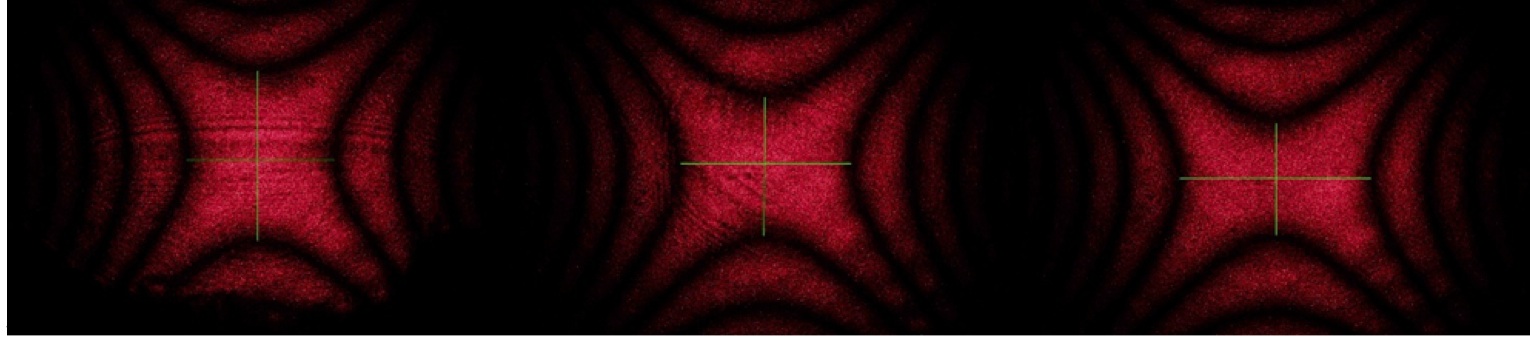}
\caption{\label{fig3} Typical fringe patterns carried out by observing in the $(a-c)$ crystallographic plane. The shape of the fringes are quartic curves which resemble hyperbolae. The distances between the hyperbole branches (the green lines in the pictures) change as a function of the stress state. These distances are easy measurable and can be linked to the stress condition by a simple model.}
\end{figure}

Two different photoelastic methods have been used in order to inspect the sample: laser Conoscopic technique \cite{17, 18}, which provides a fine spatial resolution, and laser Sphenoscopic technique \cite{23}, less detailed but faster. The two setups are similar, the only difference being the probe volume shape: a conic probe for Conoscopy and a wedge-shaped probe for Sphenoscopy. The different shape of the probe volume is obtained by a different configuration of lens system in the experimental apparatus.
These set ups are largely used to observe and analyse PWO samples along their optical axis, which corresponds to the c crystallographic axis \cite{17}-\cite{19}, \cite{23, 24}; any mechanical stress within the material warps the circular isochromates into Cassini-like curves \cite{23}.
Otherwise, by observing in $(a-c)$ planes, the isochromates resemble either hyperbolic or parabolic curves (Figure~\ref{fig3}) \cite{24, 25} and no geometrical reference in the unstressed state is recognizable (e.g. circular isochromates observed along the $c-$direction). 

From the interference pattern, easy measurable quantities have been defined in order to extract information on the stress condition and its variation. By image processing techniques, the distance between the tips of the hyperbole-like fringes is measurable with a considerable reliability (Figure~\ref{fig3} in green). That distance, experimentally, varies as a function of the sample status. Actually, the difference between the vertical and the horizontal distance seems to be a good signature of the stress condition in the inspected volume. 

\subsection{Data interpretation: the proposed elasto-optic model}

A theoretical model is proposed to interpret the complex Conoscopic fringe pattern and extract information about stress distribution. The model is based on the Bertin surfaces \cite{12}, \cite{16}, \cite{24}-\cite{26} distortion generated by the piezo-optic effect \cite{15}, \cite{25}, \cite{27}-\cite{29} which depends on the crystal piezo-optic matrix $\Pi$ and on the dielectric impermeability tensor $\mathbf B$ \cite{27}-\cite{29}. 

\begin{figure}[htbp]
\centering 
\includegraphics{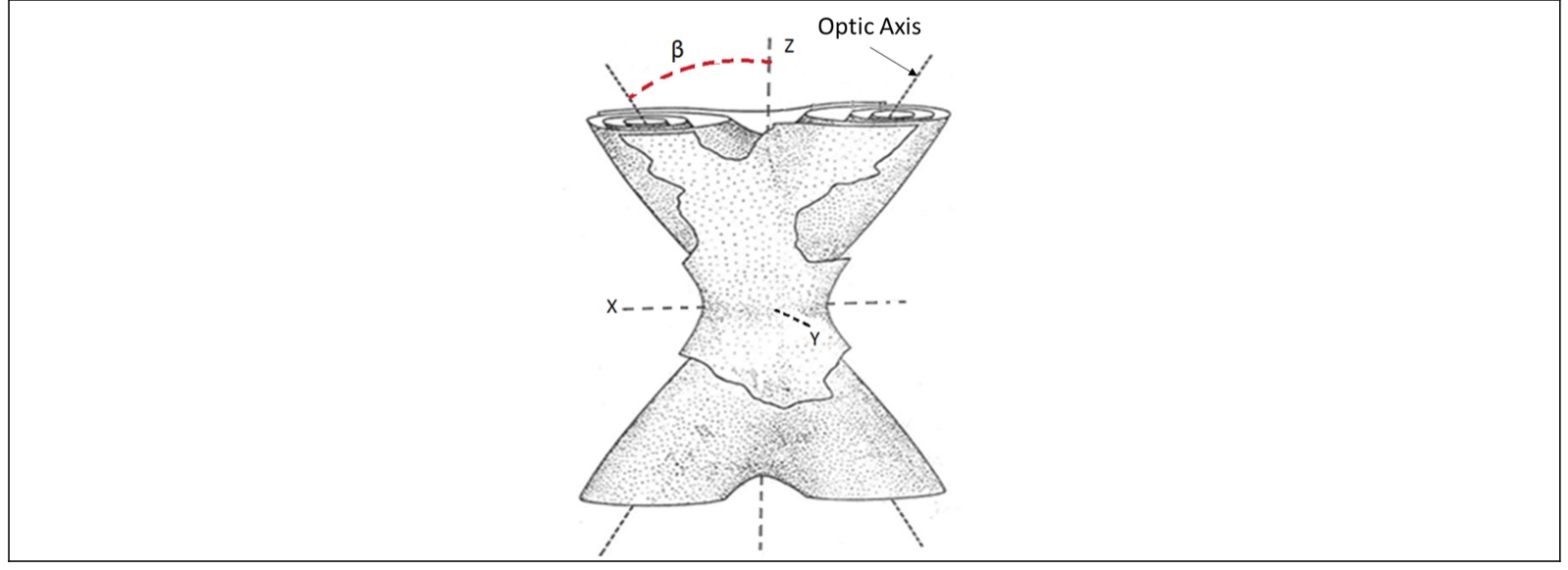}
\caption{\label{fig4} Bertin surfaces which represent the the loci of point where the light rays, crossing the sample, have the same delay. In the figure the general case of surface in a biaxial crystal is represented. Such condition which is natural in biaxial crystal, can also be induced by applying a stress to an uniaxial crystal.}
\end{figure}

Here the main steps of the model are summarized, further details about the mathematical procedure can be founf \emph{e.g.\/} into \cite{24}-\cite{26}. We identify the crystallographic frame with $c=z, a=y$ and $b=x$, taking into account that since PWO is a tetragonal crystal then  $a=b$. In the most general case, the equation of the Bertin surfaces for a biaxial crystal are \cite{16}, \cite{24}-\cite{26}:
\begin{equation}\label{bertin1}
\cos^{4}\beta x^{4}+y^{4}+\sin^{4}\beta z^{4}+2\cos^{2}\beta x^{2}y^{2}-2\sin^{2}\beta\cos^{2}\beta x^{2}z^{2}+2\sin^{2}\beta y^{2}z^{2}-H^{2}(x^{2}+y^{2}+z^{2})=0
\end{equation}
where $\beta=\beta(\mathbf B)$ is the semi-angle between the optical axes (vid. Figure~\ref{fig4}), and $H$ is defined by
\begin{equation}\label{acca}	
H=\frac{\lambda N}{n_{e}-n_{o}}\,,
\end{equation}
where $\lambda$ is the wavelength and $N$ is the fringe order. 

The relative surfaces (for each N integer value) are shown in Figure~\ref{fig4}, which represents the classical Bertin equal delay surfaces due to a biaxial condition of a crystal sample which can occur naturally (depending on the crystal system) or when a uniaxial crystal undergoes to specific load directions; the latter is the PWO case, analyzed in this work. 

\begin{figure}[htbp]
\centering 
\includegraphics{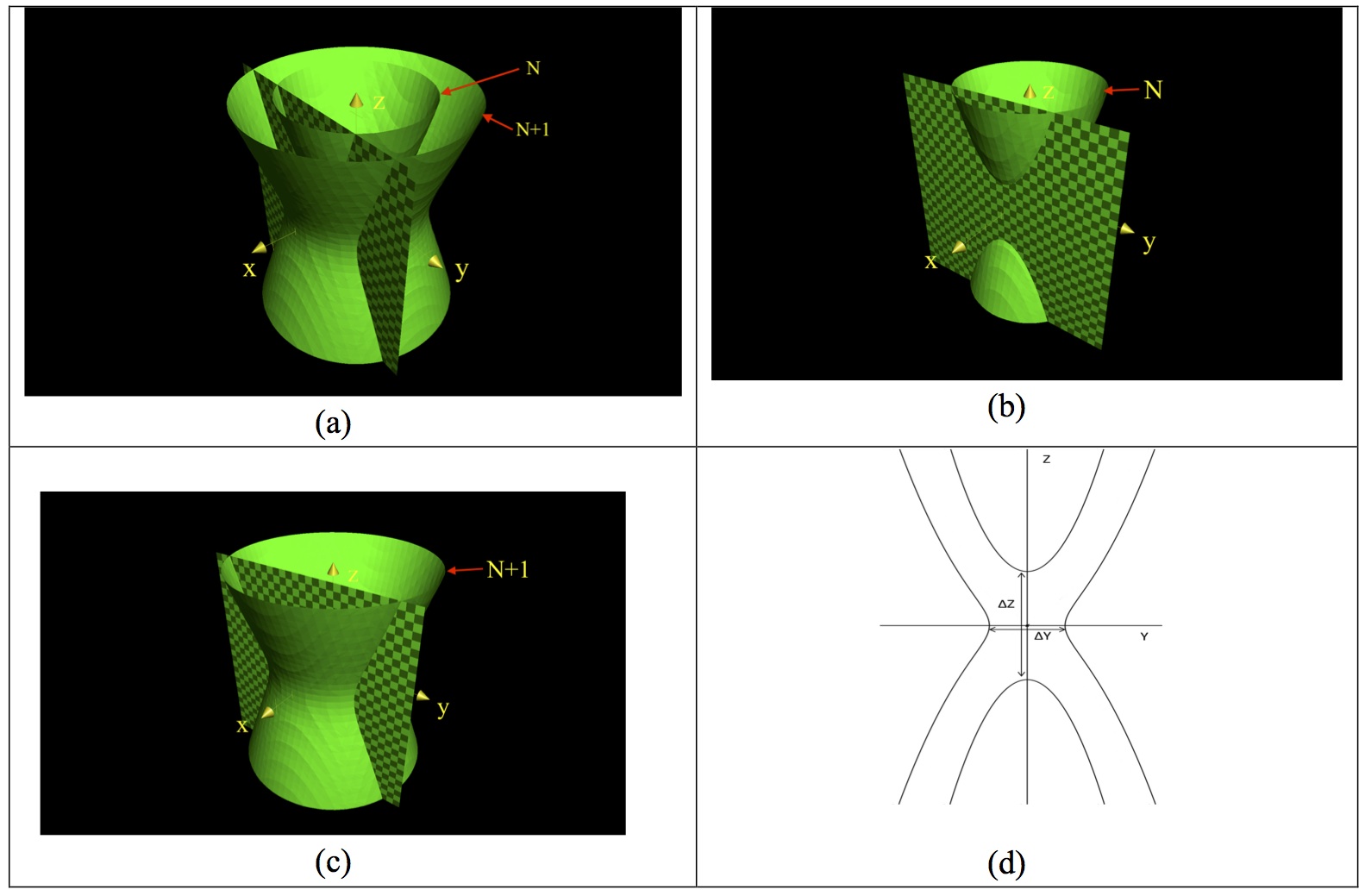}
\caption{\label{fig5} Scheme of the Bertin surfaces in uniaxial condition (a) for two orders $(N, N+1)$ and the intersection curves. For clarity in (b) we draw only the intersection with the $N$ order, and in (c) with the $N+1$ order. In (d) both the intersection curves are plotted. They represents the shape of the fringes which can be observed in the $(a-c)$ plane observation. The two measurable quantities, $\Delta Y$ and $\Delta Z$, are also indicated. Those quantities are a function of the state of stress.}
\end{figure}

In Figure~\ref{fig5} the intersection of the Bertin surfaces with the observation plane $x=d$, where $d$ is the sample thickness, and the relative interference curves $f(d\,,y\,,z)=0$ are presented. 

Actually, this intersection lead to the formation of quartic curves $g(y\,,z)=\xi$ (Figure~\ref{fig5}d)\cite{24, 25}, parameterized on subsequent fringe orders by the means of $\xi=d/NH$. As experimentally measurable quantity we assume the distances $(\Delta z\,,\Delta y)$ between the apex of the curves which result by sectioning the Bertin surfaces (Figure~\ref{fig5}), respectively with $(y=0\,,x=d)$ to obtain $\Delta z$ and  $(z=0\,,x=d)$ for $\Delta y$.

The scheme presented in Figure~\ref{fig5}, refers to the unstressed condition; due to the piezo-optic effect expressed by the fouth-order piezo-optic tensor $\Pi$, the dielectric impermeability tensor $\mathbf B$ modifies as function of the load condition: 
\begin{equation}\label{piezo}
\mathbf B(\mathbf T)=\mathbf B+\Pi[\mathbf T]\,,
\end{equation}
where $\mathbf T$ is the symmetric Cauchy stress tensor. By (\ref{piezo}) the Bertin surfaces will deform and the values of $\beta$ (Figure~\ref{fig4}), $\Delta y$ and $\Delta z$ (Figure~\ref{fig5}d) will change accordingly. Since we deal with plane photoelasticity, only plane stress condition can be detected, the only non-null components of $\mathbf T$ being denoted as $\sigma_{zz}\,,\sigma_{yy}$ and $\tau_{zy}$.

The difference $R=\Delta y-\Delta z$ then will be a function $R=R(\mathbf T\,,\xi)$; for "small" stress such a difference can be linearized to within higher-order terms in $\|\mathbf T\|$ to obtain
\begin{equation}
R(\sigma_{zz}\,,\sigma_{yy})=k_{0}(\xi)+k_{1}(\xi)\frac{\pi_{12}-\pi_{11}}{n_{o}^{-2}-n_{e}^{-2}}\sigma_{yy}+o(\sigma_{zz}\,,\sigma_{yy})\,,
\end{equation}
where $k_{0}$ and $k_{1}$ are explicit function of $\xi$ whereas $\pi_{12}\,,\pi_{11}$ are two components of $\Pi$.
\subsection{Photoelastic tests and results}

\subsubsection{Conoscopic inspection}

\begin{figure}[htbp]
\centering 
\includegraphics{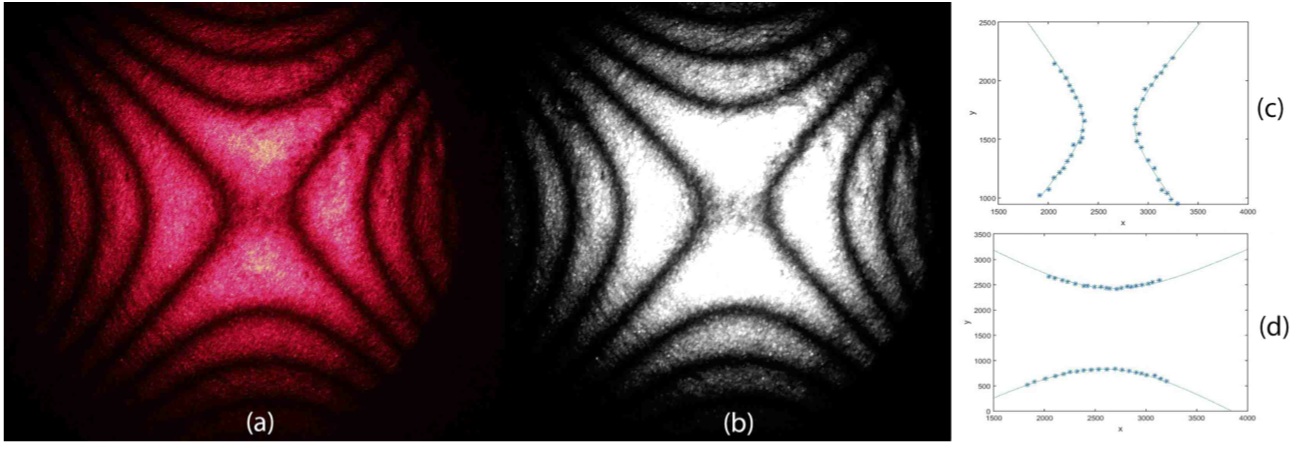}
\caption{\label{fig6} Conoscopic pattern of the isochromate fringes: (a) raw image carried out observing orthogonally to the optic axis. (b) processed image, (c) and (d) fitting of the isochromatic fringes with hyperbolae.}
\end{figure}

By Laser Conoscopy technique, a grid of points (20x20) with resolution of 1 mm, has been inspected. From each point, a fringe pattern image has been obtained. 
Observations have been carried out along the a crystallographic axis, orthogonal to the optical axis. 
The sample has been placed in position in order to observe a symmetric pattern. This last step confirmed that the a crystallographic axis of the sample is tilted of about $4^{\circ}$ with respect to the crystal surface, in fully agreement with XRD results. 
In Figure~\ref{fig6} the observed fringe pattern obtained by Conoscopy and the relative process to extract the fringe shape are shown. 

\begin{figure}[htbp]
\centering 
\includegraphics{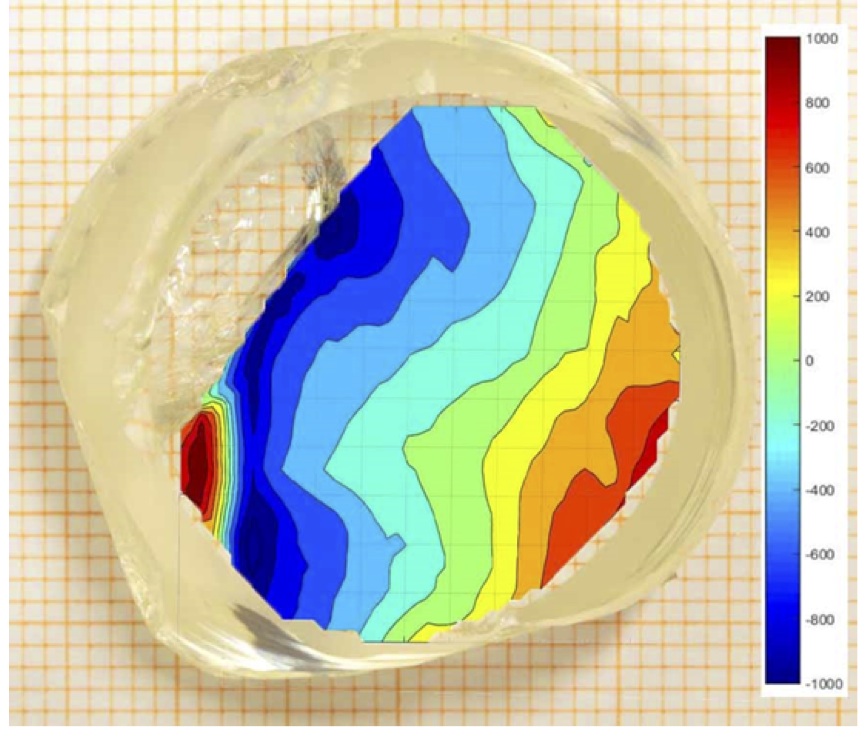}
\caption{\label{fig7} Map of $R(\sigma)=\Delta y-\Delta z$ superimposed to the sample geometry. Colours in the scale are proportional to the intensity of $R$ (with sign, measured in pixels).}
\end{figure}

By a pointwise measure of $R(\sigma)$, it is possible to obtain a map of the variation of the residual stress. Figure~\ref{fig7} shows the spatial distribution in the PWO crystal of the measured values of $R(\sigma)$. 

\subsubsection{Sphenoscopic inspection}

\begin{figure}[htbp]
\centering 
\includegraphics{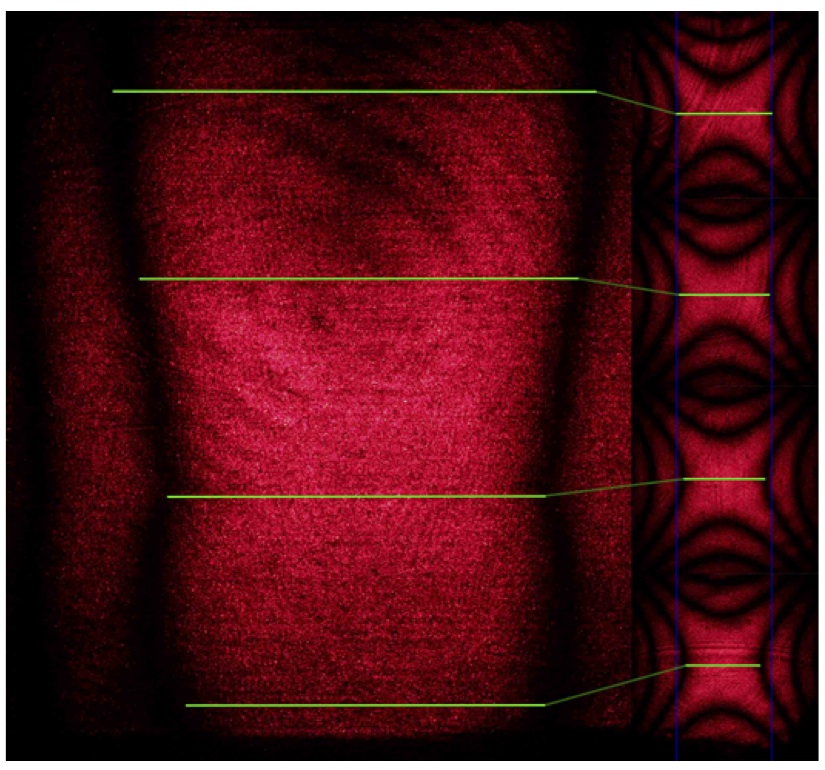}
\caption{\label{fig8} Comparison between Sphenoscopy and Conoscopy. The image on the left, is the Sphenoscopic pattern carried out close to the left boarder of the sample along the y axis. Warped interference fringes indicate the presence of stress variation along $z-$ axis. Horizontal (green) segments evidence the distance between fringes. In the right part fringes taken by Conoscopy, at the same y values, are reported. Segments represent the distances between the hyperbole branches.}
\end{figure}

The Sphenoscopic probe beam has been moved in the horizontal $(y)$ and vertical $(z)$ directions so that 40 images (20+20), consisting of Sphenoscopic fringe patterns, have been obtained. Since the light spans over a plane angle from the focal line, only one dimension of the fringe orders is generated along the height of the light wedge (Figure~\ref{fig8}). In case of absence of stress and defects in the probe volume, the fringe pattern is composed of a series of straight lines. The presence of stress or defects is indicated by variation of distance between fringes, so that fringe curvature can be associated to stress gradients along the light wedge height (Figure~\ref{fig8}).

\begin{figure}[htbp]
\centering 
\includegraphics{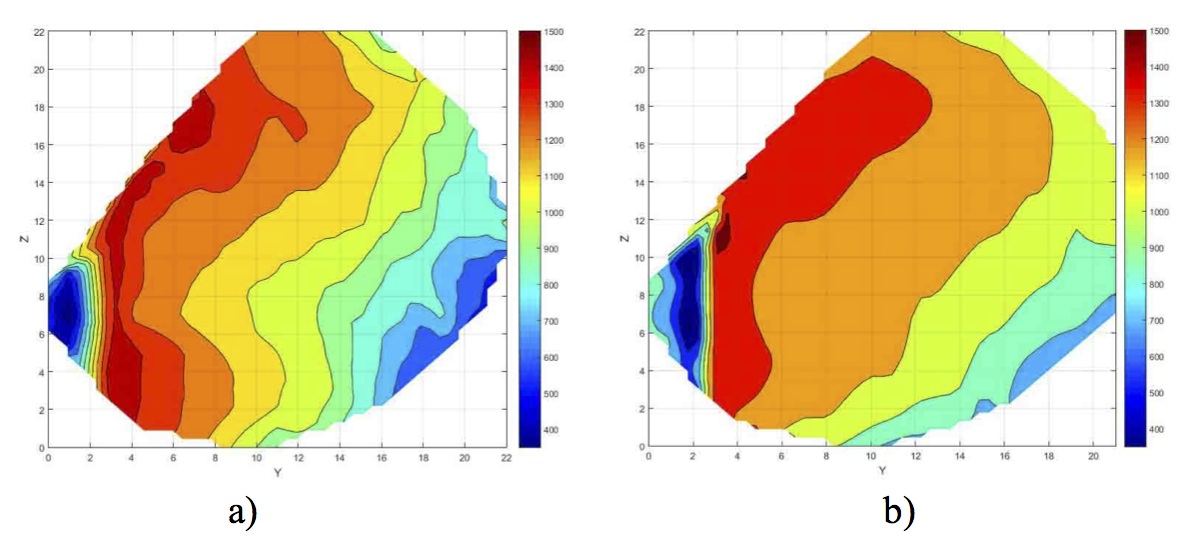}
\caption{\label{fig9} Trend of the vertical distance between fringes: a) Conoscopic results, b) Sphenoscopic results (wedge along the $y-$axis).}
\end{figure}

When the light wedge coincides with the optic axis, the fringe pattern maps the variation of the horizontal distances $(\Delta y)$ between the conoscopic fringe orders. On the other hand, when it is orthogonal to the optic axis, the fringe variation describes the vertical distances $(\Delta z)$. In Figure~\ref{fig9} the image processing sequence is shown. In the left part, the fringe distances are evaluated by means of Sphenoscopy. In the right part, they are compared to the corresponding Conoscopic distances of the hyperbole vertexes. Therefore, Figure~\ref{fig9} provides comparison between the two different techniques, which are clearly correlated to each other and sensitive to stress condition over the sample. Experimental data acquired by Sphenoscopy and Conoscopy have been coherently treated. In Figure~\ref{fig9} the Conoscopic (Figure~\ref{fig9}a) and Sphenoscopic (Figure~\ref{fig9}b) data are reported for comparison. 

Although, as expected, the Conoscopic analysis is more detailed, the Sphenoscopic map confirms the trend of the stress distribution (Figure~\ref{fig9}), and it is suitable for a faster inspection, often preferred by industry.

\section{Discussion}

A quick visual inspection of the sample gives an idea of the defectiveness consisting in lack of geometrical symmetry, scratched and fractured surface. 

However, in spite of the visible defective shape of the sample, results of the XRD investigation clearly show that the analysed crystal has a tetragonal structure with lattice parameters only slightly higher than the nominal PbWO$_4$ (stolzite) compound (Figure~\ref{fig2}). The difference between experimental and nominal values of the lattice parameters can be attributed to the higher content of Pb and W in the crystal, as evidenced by EDS analysis (Table~\ref{tab1}). In fact, higher content of Pb and W, that are elements with higher atomic radius with respect to O, causes an increase of the crystallographic cell with a consequent increase of the lattice parameters.

Furthermore, XRD investigations carried out on the AP sample evidenced a preferential orientation growth of the crystal with the $(2,0,0)$ lattice planes parallel to the sample surface (Figure~\ref{fig2}). This latter result is further confirmed by rocking curve, which allows also estimating a shift of about $4^{\circ}$ with respect to the surface normal (inset in Figure~\ref{fig2}). It is worth to note that photoelastic measurements confirmed the same shift observed by XRD.

Photoelastic techniques allowed obtaining a qualitative map of residual stress distribution that is not predictable only by a thermal stress simulation \cite{30, 31}. The fringe pattern distortion over the sample, observed by Conoscopy and/or Sphenoscopy, has been correlated to the residual stress distribution inside the crystal via a simple theoretical model capable of interpreting the shape of fringes in terms of residual stress.

The elasto-optic model links variations of the measurable quantity $R(\sigma)=\Delta z-\Delta y$ to the stress distribution over the sample.

Generally, $R(\sigma)$ is a complicated function of the stress components $\sigma_{yy}$ and $\sigma_{zz}$. Nevertheless, under hypotheses of ''small stress'' it can be represented as a linear function of $\sigma_{yy}$; the $\sigma_{zz}$ stress component become significant at higher orders (see the Appendix).

Actually, by the anisotropy of PWO in a linear approximation the $\sigma_{zz}$ stress component cannot be detected at the first order, a fact that must be taken into account from both a theoretical and technological point of view.

Despite we cannot measure the stress magnitude (by lacking, at the moment, of knowledge of piezo-optic tensor components for PWO, in fact the evaluation of the piezo-optic $\Pi$ tensor, \emph{see\/} the Appendix, is quite complex and it has been completely calculated for few crystals \cite{32}-\cite{34}), the gradient of the measured quantity $R(\sigma)$ describes the variations of the residual stress within the sample. In fact, great evidence of such variation can be recognized from changes of distances between hyperbolic fringes. Whatever the load combination is, the distance between the hyperboles branches changes whenever a variation of stress is present. The normalized load map is powerful tool to describe the stress gradient and status of materials, especially when an exact magnitude cannot be calculated. This method is largely used in material science \cite{15}, \cite{35},  giving a lot of useful information on crystal conditions.

Despite the complexity of the phenomenon and the effectiveness of the analyzed sample, a map of stress distribution has been produced by this method, which shows a good reliability. In fact, although a further calibration step would be necessary to achieve quantitative information on stress, the obtained map well represents gradient of residual stress which can be easily recognized close to scratches and near the border (Figure~\ref{fig7}). The map presents higher gradient of the $R(\sigma)$ next to the highly defective zones, where chipping, scratches and fractures are clearly visible, while smoother gradients are detected on the opposite side where the sample seems to be less defective.
It is also easy to see that the gradient flow has the same direction of the ''sliding'' phenomena that affects the distorted cylinder, while small variations are recognizable in the orthogonal $z$ direction. 

It is worth to remark that in case of good quality of the sample, no variation of fringe distances should be recognizable and a flat map should be expected by Conoscopic and Sphenoscopic inspections.

\section{Conclusions}

A complete characterization of a PWO sample cut and observed in the plane containing the optic axis $(a-c)$ have been performed. Sample originates from a defective crystal, which has been rejected from the production line. XRD has shown that the $(2,0,0)$ lattice planes grow parallel with an angle of $4^{\circ}$ with respect to the sample surface. EDS microanalysis evidences absence of any dopant and contaminant as well as a higher content of Pb and W with respect to the nominal stoichiometry of the PbWO$_4$ compound. This higher concentration of elements causes an increase of the lattice parameters of the tetragonal crystal structure.

The photoelastic methods have confirmed the $4^{\circ}$ misalignment detected by the XRD. For the photoelastic inspection a novel method has been proposed in order to carry out the stress analysis in the $a-c$ plane. The proposed elasto-optic model, aimed at interpreting the data from the photoelastic investigations, provides some insight into the phenomena occurring in the crystal and allows for a reliable analysis of the homogeneity of the stress distribution.

The reliability of the method is validated by the results, which are in agreement with the crystal defectiveness, inhomogeneity of the condition and anisotropy. The agreement between the Conoscopy and Sphenoscopy is good in terms of spatial distribution. However, Conoscopy provides a better spatial resolution than Sphenoscopy, at the expense of complexity and measurement time required for the scanning process. 

It is remarkable that the theoretical results are applicable to all tetragonal crystals of the point group $4/m$, but the theoretical approach and the methodology is exploitable for all the crystal classes.

\acknowledgments

This work has been developed within the framework of the Crystal Clear Collaboration (CERN R\&D Experiment 18) and with the partial support of the COST Action TD1401-Fast Advanced Scintillator Timing (FAST). It is also supported by the Departments SIMAU, DIISM and DICEA of the Universit\'a Politecnica delle Marche. Crystal sample has been provided by Crytur (CZ). We wish to thank a reviewer for his useful comments and Michel Lebeau, former CERN associate, for his support to our research work.

\end{document}